\documentclass[aps,prl,reprint,groupedaddress,longbibliography]{revtex4-1}

\usepackage{graphicx}
\usepackage{amsmath}
\usepackage{amssymb}
\usepackage{longtable}

\newcommand{\OOp}{O$_2^+$} 
\newcommand{\p}{\ensuremath{\prime}}
\newcommand{\pp}{\ensuremath{{\prime\prime}}} 
\newcommand{\ket}[1]{\ensuremath{|#1\rangle}}

\begin{document}

\title{High sensitivity to mass-ratio variation in deep molecular potentials}

\author{D. Hanneke}
\email[]{dhanneke@amherst.edu}
\author{R. A. Carollo}
\author{D. A. Lane}
\affiliation{Physics \& Astronomy Department, Amherst College, Amherst, Massachusetts 01002, USA}

\date{\today}

\begin{abstract}
Molecular vibrational transitions are prime candidates for model-independent searches for variation of the proton-to-electron mass ratio. Searches for present-day variation achieve highest sensitivity with deep molecular potentials. We identify several high-sensitivity transitions in the deeply bound O$_2^+$ molecular ion. These transitions are electric-dipole forbidden and thus have narrow linewidths. The most sensitive transitions take advantage of an accidental degeneracy between vibrational states in different electronic potentials. We suggest experimentally feasible routes to a measurement with uncertainty exceeding current limits on present-day variation in $m_p/m_e$.
\end{abstract}


\maketitle

The dimensionless fundamental constants are the input parameters to our physical theories. Apparent variations of these constants arise naturally in many extensions to the Standard Model, including the spacetime evolution of additional dimensions 
or new scalar fields~\cite{uzanRMP2003}.
Recent work suggests that certain dark matter fields could induce oscillations in the values of fundamental constants~\cite{stadnikPRL2015}.

The proton-to-electron mass ratio, $\mu=m_p/m_e$, is particularly interesting because the two masses arise from different mechanisms. Variation would imply a change in the relative strengths of the strong and electroweak interactions. Models typically predict the relative change of $\mu$ should be of order 40 times larger than that of the fine structure constant $\alpha$~\cite{uzanRMP2003}.

Searches for variation of $\mu$ have been approached over both cosmological and laboratory timescales. The current precision of cosmological searches are at the level of $10^{-6}$--$10^{-7}$ over $\sim10^{10}$ years~\cite{kanekarApJL2011,bagdonaitePRL2013,ubachsRMP2016}.
The tightest bounds on present-day variation of $\mu$ come from atomic clock experiments, which set the limit $\dot{\mu}/\mu\lesssim 10^{-16}~{\rm yr}^{-1}$~\cite{godunPRL2014,huntemannPRL2014}.
In these experiments, nearly all the sensitivity to $\mu$ variation comes from the hyperfine structure of cesium, and extracting the precise $\mu$ dependence requires a model of the cesium nuclear magnetic moment~\cite{flambaumPRC2006}.

The vibration and rotation of molecules provide a model-independent means to search for variation in $\mu$~\cite{carrNJP2009,chinNJP2009,jansenJCP2014}.
The most stringent constraint from a molecular measurement is $\dot{\mu}/\mu = (-3.8\pm 5.6)\times10^{-14}~{\rm yr}^{-1}$ in SF$_6$~\cite{shelkovnikovPRL2008}.
We propose \OOp~as a molecule possessing a high sensitivity to present-day variation in $\mu$ as well as experimentally feasible means for measuring it. We describe two possible measurements, each of which is capable of resolving fractional changes in $\mu$ to better than $10^{-16}$ in one day with a single molecule. As discussed below, the high sensitivity arises from the molecule's deep electronic ground-state potential (54\,600~cm$^{-1}$). Other molecules with deep potentials may also have suitable transitions.

Features of the relatively simple molecular structure of \OOp~make it amenable for experiments. It is homonuclear, so nuclear symmetry eliminates half the rotational states and forbids electric dipole (E1) transitions within an electronic state. This nonpolarity suppresses many systematic effects, including some AC Stark and blackbody radiation shifts. The most common isotope of oxygen ($^{16}$O, 99.8\,\% abundance) has no nuclear spin, so \OOp~lacks hyperfine structure. Unlike many molecular ions, \OOp~has measured spectroscopic parameters~\cite{krupenieJPCRD1972, cosbyJMS1980, hansenJMS1983, coxonJMS1984, kongCJP1994, kongIJMSIP1996, songJCP1999, songJCP2000, songJCP2000a}
~and existing theoretical calculations~\cite{fedorovJCP1999, fedorovJCP2001, zhangMolPhys2011, magrakvelidzePRA2012, liuMolPhys2015}. This prior work has been motivated in part because of the important role \OOp~plays in the ionospheres of Earth and other planets~\cite{schunkRevGeophysSpacePhys1980}. Most relevant to the present work, several vibrational states in the middle of the \OOp~ground~$X\,^2\Pi_g$  potential are nearly degenerate with low vibrational states of the excited $a\,^4\Pi_u$ potential. This degeneracy should allow searches for variation in $\mu$ with high sensitivity in both the absolute and relative senses~\cite{demillePRL2008}.

Searches for fractional changes in $\mu$ usually involve monitoring the energy difference $\hbar\omega$ between two energies with different $\mu$-dependence, $\hbar\omega=E^\prime(\mu)-E^{\prime\prime}(\mu)$. The change in $\mu$ is then given by
\begin{equation}
	\frac{\Delta\mu}{\mu} = \frac{1}{\mu}\frac{\partial\mu}{\partial\omega}\Delta\omega = \frac{\partial(\ln\mu)}{\partial\omega}\Delta\omega.
\end{equation}
The absolute sensitivity of the transition is given by $\partial\omega/\partial(\ln\mu)$, which is sometimes called the absolute enhancement factor. In an experiment, the statistical precision with which $\Delta\omega$ can be measured is given by
\begin{equation}
	\delta\omega = \frac{\Gamma}{\sqrt{M}\,S/\delta S}, \label{eq:statistics}
\end{equation}
where $\Gamma$ is the transition linewidth, $S/\delta S$ is the signal-to-noise ratio, and $M$ the number of independent measurements (assuming white noise). Here, $\delta\omega$ represents the uncertainty in determining the change $\Delta\omega$.
The figure of merit is thus 
\begin{equation}
 \frac{\partial\omega}{\partial(\ln\mu)}\frac{1}{\Gamma}.
\end{equation}
In some cases, such as the Doppler-broadened lines encountered in astrophysical measurements, the linewidth is proportional to the transition frequency and the figure of merit is proportional to the relative enhancement factor $[\partial\omega/\partial(\ln\mu)]/\omega$. In other cases, such as \OOp, such relative enhancement can be experimentally convenient.

Because of its importance in isotope shifts, the scaling of molecular parameters with $\mu$ has been known for some time~\cite[Sec. III.2.g]{herzbergVolI1950}. In particular, for a state of energy
\begin{equation}
	E/(hc) = T_e + \omega_e (v+\tfrac{1}{2}) - \omega_ex_e (v+\tfrac{1}{2})^2 + B_e J(J+1),
\end{equation}
the electronic energy $T_e$ is independent of $\mu$, the vibrational coefficient $\omega_e$ scales as $\mu^{-1/2}$, the lowest anharmonicity coefficient $\omega_ex_e$ scales as $\mu^{-1}$, and the rotational constant $B_e$ scales as $\mu^{-1}$. (Here, the parameters are given as wavenumbers. For scaling of additional coefficients, see references~\cite{herzbergVolI1950,beloyPRA2011,pastekaPRA2015}.) Thus the absolute sensitivity of a particular state to variation in $\mu$ is given by
\begin{equation}
	\frac{1}{hc}\frac{\partial E}{\partial(\ln\mu)} = -\tfrac{1}{2}\omega_e(v+\tfrac{1}{2}) + \omega_ex_e (v+\tfrac{1}{2})^2 - B_e J(J+1). \label{eq:Esens}
\end{equation}
Transitions between different vibrational states will generally yield higher sensitivity both because $\omega_e$ tends to be larger than $B_e$ and because selection rules preclude transitions between states of vastly different $J$. The first term in Eq.~(\ref{eq:Esens}) shows a linear growth in sensitivity with vibrational state. For higher states, the opposite sign of the second term slows the growth. The vibrational states return to no sensitivity near the dissociation limit. As was pointed out in refs.~\cite{zelevinskyPRL2008,demillePRL2008}, the peak sensitivity is approximately 1/4 of the dissociation energy and occurs for vibrational states with energies approximately 3/4 of the dissociation limit.

Vibrational selection rules typically preclude direct transitions between low- and high-sensitivity states within the same electronic state. To alleviate this restriction, Zelevinsky~\emph{et al.}~\cite{zelevinskyPRL2008} proposed driving stimulated Raman transitions via an excited electronic state and suggested Sr$_2$ as a candidate molecule. DeMille~\emph{et al.}~\cite{demillePRL2008} suggested transitions between different electronic states. The linewidth for such a transition can still be narrow if the inter-electronic transition is forbidden, for example by spin selection rules. DeMille~\emph{et al.} emphasize the practical advantage of choosing transitions with high relative sensitivity and identifies Cs$_2$ as a candidate molecule with a near-degeneracy between vibrational states in different-multiplicity electronic states.

Because the maximum sensitivity is proportional to the potential depth, one should look for experimentally viable routes in deeply bound molecules. We have identified \OOp~as a candidate molecule with several accessible transitions that are 50--75 times more sensitive than those in prior proposals with photoassociated molecules. 
Indeed, even the energy difference between the \OOp~$X\,^2\Pi_g$ ground and first-excited vibrational states has several times the absolute sensitivity of the transitions proposed in refs.~\cite{zelevinskyPRL2008,demillePRL2008}. Additionally, there are accidental degeneracies between the 21$^{\rm st}$ and 22$^{\rm nd}$ excited vibrational states of the $X\,^2\Pi_g$ state and $v = 0, 1$ of the $a\,^4\Pi_u$ state. Several transitions between these states are likely to have energies in the microwave range. Spin-orbit coupling between $a\,^4\Pi_u$ and the nearby $A\,^2\Pi_u$ state should allow the driving of these nominally spin-forbidden transitions~\cite{LefebvreBrionAndField}.

\begin{figure}
	\includegraphics[width=\columnwidth]{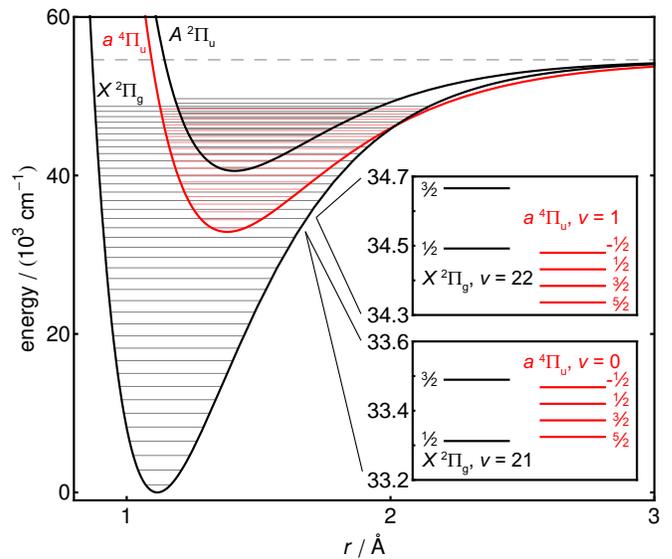}%
	\caption{Potential curves (in the Morse approximation) for the $X$, $a$, and $A$ states of \OOp. The horizontal lines indicate the measured energies of vibrational states~\cite{songJCP1999,songJCP2000,songJCP2000a}. Inset are the doublet-$X$ and quartet-$a$ levels discussed in the text, including spin--orbit splittings. The labels on each fine-structure level indicate $\Omega$ in the case (a) (low-$J$) limit.\label{fig:pot}}
\end{figure}

\begin{figure}
	\includegraphics[width=\columnwidth]{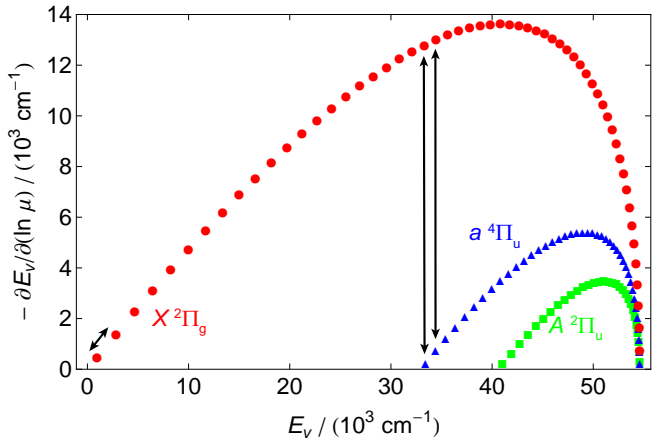}%
	\caption{Absolute sensitivity of vibrational states in the $X$, $a$, and $A$ potentials, calculated using the Morse approximation. The arrows indicate the proposed transitions.\label{fig:sens}}
\end{figure}

The lowest molecular potentials of \OOp~have been studied for some time. Figure~\ref{fig:pot} plots the $X\,^2\Pi_g$, $a\,^4\Pi_u$, and $A\,^2\Pi_u$ potentials. The vibrational state energies have been measured up to $v=38$ for the $X$ state~\cite{kongCJP1994,songJCP1999}, $v=18$ for the $a$ state~\cite{kongIJMSIP1996,songJCP2000a}, and $v=12$ for the $A$ state~\cite{songJCP2000}. By use of the resulting molecular parameters as well as Eq.~(\ref{eq:Esens}), we calculate each vibrational level's sensitivity to variation in $\mu$. These sensitivities, $\partial E_v/\partial(\ln\mu)$, are plotted in Fig.~\ref{fig:sens}. The values plotted in the figure are calculated using a Morse approximation for the potential~\cite{herzbergVolI1950}. For the particular transitions proposed herein, the sensitivity calculated from the Morse potential and from the measured molecular parameters agree to better than 0.5~\%.

The $X$-state's high $\omega_e$ means that even the lowest vibrational transitions are quite sensitive to variation in $\mu$. The transition $|X, v=0\rangle\leftrightarrow |X,v=1\rangle$, has a sensitivity of $\tfrac{1}{2\pi c}\tfrac{\partial\omega}{\partial(\ln\mu)} = 920~{\rm cm}^{-1} = 28~{\rm THz}/c$ and an energy difference $\Delta E/(hc) = 1873~{\rm cm}^{-1} = 1/(5339~{\rm nm})$. Because \OOp~is nonpolar, this transition is E1 forbidden but proceeds as an electric quadrupole (E2) transition. Its natural linewidth is thus extremely narrow and any experimental linewidth will be limited by technical noise (e.g. laser linewidth) or probe time. An experiment driving the lowest vibrational transitions with two Raman lasers has been proposed in N$_2^+$~\cite{kajitaPRA2014}. The N$_2^+$ ground-state $v=0\leftrightarrow 1$ electric-quadrupole transition has been driven directly with a quantum cascade laser~\cite{germannNaturePhys2014}. Similar techniques could be applied to \OOp.

Given the absolute sensitivity, we can use Eq.~(\ref{eq:statistics}) to estimate the achievable statistical precision of a $v=0\leftrightarrow1$ measurement. Assuming a probe time equal to $\Gamma^{-1}$ and minimal experimental dead time, the total number of measurements scales linearly in the total measurement duration $\tau$ as $M=\tau\Gamma$. If the signal-to-noise is limited by quantum projection noise~\cite{itanoPRA1993}, then $S/\delta S = \sqrt{N}$, where $N$ is the number of independent molecules probed per experimental run. The statistical precision would then be $\delta\omega\sim\sqrt{\Gamma/(N\tau)}$. With a $\Gamma/(2\pi)=1~{\rm Hz}$ linewidth, the lowest vibrational transition should be able to achieve $\delta\mu/\mu \sim 1.4\times10^{-14}/\sqrt{N (\tau/{\rm sec})}$ or of order $5\times10^{-17}$ in one day with one molecule.

To enhance sensitivity, one could measure the energy difference between vibrational states near the middle of the potential and those near the bottom or near dissociation. With a potential as deep as \OOp, driving such a transition with two Raman lasers becomes challenging. Directly driving the quadrupole overtone transitions suffers from very small quadrupole moments for large $\Delta v$. In \OOp, accidental degeneracies between different electronic potentials provide high sensitivity with relatively low energy difference. Here, two high-sensitivity states $|X\,^2\Pi_g, v=21,22\rangle$ are nearly degenerate with two low-sensitivity states $|a\,^4\Pi_u, v=0,1\rangle$. Figure~\ref{fig:pot}(inset) shows the overlap, including spin-orbit splitting. Because the rotational coefficients of these two states are slightly different, the degeneracy may in some sense be ``tuned'' by choosing an appropriate $J$ and $\Delta J$. The absolute sensitivity of the $\ket{X,v^\pp=21}\leftrightarrow\ket{a,v^\p=0}$ transition is $12\,600~{\rm cm}^{-1}=378~{\rm THz}/c$; that of the $\ket{X,v^\pp=22}\leftrightarrow\ket{a,v^\p=1}$ transition is $12\,300~{\rm cm}^{-1}=369~{\rm THz}/c$. Depending on the particular $J$ and $\Delta J$, sensitivity contributions from the $B_e$ coefficient may be of order $100~{\rm cm}^{-1}$.

\begin{figure}
	\includegraphics[width=\columnwidth]{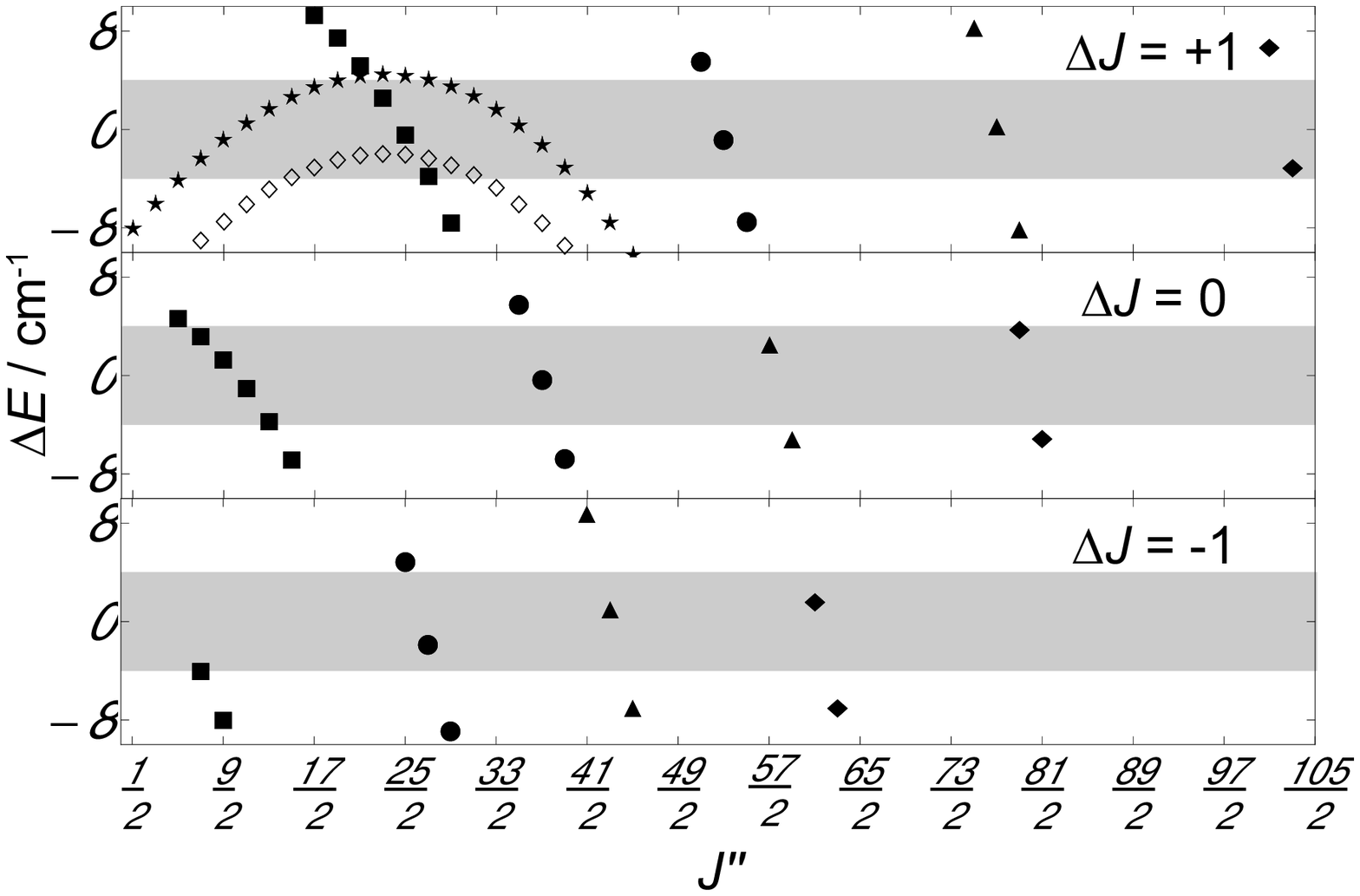}%
	\caption{Degeneracy of $X$ and $a$ states. Transitions are plotted from $\ket{X\,F_2 ,v^{\pp}=21,J^{\pp}}$ (open) or $\ket{X\,F_1,v^{\pp}=21,J^{\pp}}$ (filled) to $\ket{a\,F_1, v^\p=0,J^\p}$ ($\blacksquare$), $F_2$ ({\Large${\bullet}$}), $F_3$ ({\large$\blacktriangle$}), or $F_4$
 (\protect\rotatebox[origin=c]{45}{$\blacksquare$}). 
 The $\ket{X\,F_1, v^{\pp}=22,J^{\pp}}\leftrightarrow\ket{a\,F_4, v^\p=1,J^\p}$ transitions are plotted with $\bigstar$. The separate plots indicate $\Delta J=J^\pp-J^\p=-1$, $0$, or $+1$. Gray bands show a $\pm4~{\rm cm}^{-1}$ uncertainty range.\label{fig:degen}}
\end{figure}

Using measured molecular parameters for the $X$ and $a$ states~\cite{hansenJMS1983,kongCJP1994,songJCP1999,kongIJMSIP1996}, we make an effective Hamiltonian~\cite{brown_carrington} and calculate the energies and eigenstates of the individual $J$ states within the relevant vibrational states~\cite{hannekeTheorySupplement2016}. We calculate all transition energies with $\Delta J = 0, \pm 1$ and $|\Delta E|/(hc) < 10~{\rm cm}^{-1} = 300~{\rm GHz}/c$. Figure~\ref{fig:degen} plots the results, which are tabulated in the Supplemental Material~\cite{hannekeTheorySupplement2016}. As can be seen, many energies lie in a range where radiofrequency and microwave techniques may be used. The relatively lower transition frequencies relax the demands on relative accuracy while maintaining high absolute sensitivity. The uncertainties on the calculated transition energies are 3--5~cm$^{-1}$, but they are highly correlated such that even if these particular transitions are no longer within 10~cm$^{-1}$, others will be.

While transitions between the doublet-$X$ and quartet-$a$ states are spin-forbidden, spin-orbit mixing of the $a\,^4\Pi_u$ and $A\,^2\Pi_u$ states ($7\,625~{\rm cm}^{-1}$ apart) provides sufficient coupling. This mixing also dominates the decay of the $a$ state and thus the linewidth of our proposed transitions. With an estimate of the mixing and the known 690~ns lifetime of the $A$ state~\cite{kuoJCP1990}, we can calculate the linewidth of each transition. Only the $a\,^4\Pi_{1/2,u}$ and $a\,^4\Pi_{3/2,u}$ substates couple to the $A$ state, so we use our effective Hamiltonian to calculate the projection of each eigenstate in the Hund's case (a) basis. A similar technique was used in ref.~\cite{kuoJCP1990} to explain $a$-state-lifetime data. They used $72~{\rm cm}^{-1}$ as an \emph{ab-initio}-calculated estimate of the matrix elements $\langle a\,^4\Pi_{1/2,u}|H_{\rm SO}|A\,^2\Pi_{1/2,u}\rangle$ and $\langle a\,^4\Pi_{3/2,u}|H_{\rm SO}|A\,^2\Pi_{3/2,u}\rangle$, and we do so as well. (Ref.~\cite{minaevOptikaISpek1996} calculates a similar value for these matrix elements.) The transition linewidths fall in the range $\Gamma/(2\pi) = 0.07\textnormal{--}10~{\rm Hz}$~\cite{hannekeTheorySupplement2016}. With the same assumptions as before, a 1-Hz-linewidth transition should be able to achieve a statistical precision of $\delta\mu/\mu\sim 1.1\times10^{-15}/\sqrt{N (\tau/{\rm sec})}$ or of order $4\times10^{-18}$ in one day with one molecule.

When estimating the transition dipole moment, the same mixing of $a$ and $A$ states and spin-orbit matrix elements apply. Because the $a$ and $A$ states have similar equilibrium bond lengths, the coupling of $\ket{X,v^\pp}$ to $\ket{a,v^\p}$ is primarily through a single vibrational state $\ket{A,v^\p}$. By use of RKR potential curves generated from the data in refs.~\cite{songJCP1999,songJCP2000}, we calculate~\cite{leroyrkr21,*leroylevel82} the Franck--Condon factor between $\ket{X,v^\pp=21}$ and $\ket{A,v^\p=0}$ to be $1.8\times10^{-6}$. This value agrees with a prior published value~\cite{gilmoreJPCRD1992} that relied on older spectroscopy data to within 15\,\%. 
The electronic transition moment between $\ket{X,v^\pp=21}$ and $\ket{A,v^\p=0}$ has been calculated to be $0.503\,e a_0$~\cite{gilmoreJPCRD1992}. Combining these elements with our case (a) eigenstates, we estimate the transition dipole moment of these transitions to be of order $10^{-6}~e a_0$. A typical transition
could be driven with a Rabi frequency approximately the same as its linewidth by use of a microwave electric field of order 10--100~V/m.

Straightforward techniques exist for producing and analyzing the states of \OOp. Rovibrationally selected \OOp~molecules have been produced in the $X\,^2\Pi_g$ state with $v=0,1$, and 
$J$ up to $\tfrac{51}{2}$~\cite{dochainEPJConf2015}. The production is via resonance-enhanced multiphoton ionization (REMPI) with the selection coming from use of the $d\,^1\Pi_g$ Rydberg state in neutral O$_2$~\cite{surJCP1991}. Excitation to the Rydberg state requires two photons in the range 296.5--303.5~nm, and ionization requires a third photon, which could be at the same wavelength. Transfer from $\ket{X,v=0}$ to $\ket{a,v=0}$ could be driven coherently with a laser of wavelength 308~nm. This transition is allowed through the same $a$--$A$ spin-orbit mixing. The $\ket{X,v=0}\leftrightarrow\ket{A,v=0}$ transition has electric dipole moment $0.192\,ea_0$~\cite{gilmoreJPCRD1992} and Franck--Condon factor $1.7\times10^{-6}$. 
 A 1~mW laser focused to $50~\mu$m (intensity $2.5\times10^5~{\rm W/m}^2$) should produce a Rabi frequency of order 100~Hz.

Detection of the state could be done by driving from $a\,^4\Pi_u$ to $b\,^4\Sigma_g^-$, which has predissociation states at higher vibrational levels~\cite{hansenJCP1982}. Any population in the $\ket{X,v=21}$ state would not be transfered to the $b$ state. While preliminary measurements could take place in a beam, trapping \OOp~and sympathetic cooling to a Coulomb crystal with co-trapped atomic ions would allow longer probe times and eliminate first-order Doppler shifts. Trapping only a few atoms and molecules could enable non-destructive detection via quantum logic spectroscopy~\cite{schmidtScience2005,wolfNature2016}. Such detection could increase the duty-cycle by reducing the need to reload ions and would reduce systematic effects associated with micromotion in a radiofrequency trap~\cite{berkelandJAP1998}, though it may reduce the statistical limit because fewer molecules would be probed per experiment.

In conclusion, we have identified two routes in the \OOp~molecular ion to a high-sensitivity search for present-day variation in the proton-to-electron mass ratio. The highest sensitivity comes from an accidental degeneracy between excited vibrational levels of the $X$ state and the lowest vibrational levels of the $a$ state. We note that there is another set of degeneracies among the $\ket{X, v= 27\textnormal{--}30}$, $\ket{a, v=7\textnormal{--}10}$, and $\ket{A, v=0\textnormal{--}2}$ states~\cite{hannekeTheorySupplement2016}. The direct overlap with the $A$ state would require a more extensive linewidth calculation than described here. It is also likely that such degeneracies exist in other molecules. Some homonuclear molecules with deep electronic-ground-state potentials and different-multiplicity potentials dipping within them include Te$_2$~\cite{demilleKozlovPrivateComm}, Br$_2$, Ge$_2$, and I$_2^+$~\cite{balasubramanianChemRev1990}. The heavier of these tend to have smaller vibrational splittings, which increase the likelihood of a degeneracy. It is possible that similar transitions can be found among the infrared-inactive vibrational modes of deeply bound nonpolar polyatomic molecules.

\begin{acknowledgments}
 This material is based upon work supported by the NSF under Grant CAREER PHY-1255170 and the Research Corporation for Science Advancement.
\end{acknowledgments}

%


\clearpage
\widetext

\thispagestyle{empty}

\begin{center}
\textbf{\large Supplemental Material: High sensitivity to mass-ratio variation\\ in deep molecular potentials}\linebreak
\linebreak
D. Hanneke, R. A. Carollo, and D. A. Lane\\
\emph{Physics \& Astronomy Department, Amherst College, Amherst, Massachusetts 01002, USA}
\end{center}

\setcounter{equation}{0}
\setcounter{figure}{0}
\setcounter{table}{0}
\setcounter{page}{1}

\renewcommand{\theequation}{S\arabic{equation}}
\renewcommand{\thefigure}{S\arabic{figure}}
\renewcommand{\thetable}{S\Roman{table}}
\renewcommand{\bibnumfmt}[1]{[S#1]}
\renewcommand{\citenumfont}[1]{S#1}

\setlength\LTleft{0pt}
\setlength\LTright{0pt}
\setlength\LTcapwidth{0.9\textwidth}

\newcommand\Tstrut{\rule{0pt}{2.6ex}}
\newcommand\Bstrut{\rule[-1.5ex]{0pt}{0pt}}

\section{Effective Hamiltonian}

To calculate the energies and eigenstates of the rotational levels in a particular vibrational state, we diagonalize an effective Hamiltonian. See, for example, ref.~\cite[Eq.~10.114--10.115]{Sbrown_carrington}. We include the electronic and vibrational state energy $T_v$, spin--orbit coupling $A_v$, and rigid-body rotation $B_v$. As discussed below, higher-order terms such as centrifugal distortion $D_v$ or $\Lambda$-doubling are not necessary at our precision.

Eigenstates in both $X\,^2\Pi_g$ and $a\,^4\Pi_u$ are written in the Hund's case-(a) basis:
\begin{align}
	&c_{3/2}\left|^2\Pi_{3/2}\right\rangle + c_{1/2}\left|^2\Pi_{1/2}\right\rangle \\
	c_{5/2}\left|^4\Pi_{5/2}\right\rangle + &c_{3/2}\left|^4\Pi_{3/2}\right\rangle + c_{1/2}\left|^4\Pi_{1/2}\right\rangle + c_{-1/2}\left|^4\Pi_{-1/2}\right\rangle .
\end{align}
In these bases, the effective Hamiltonians are given by
\begin{equation}
	H(^2\Pi) = \begin{pmatrix}
	T_v + \tfrac{A_v}{2} + B_v\left[J(J+1)-\tfrac{7}{4}\right]	&	-B_v\sqrt{J(J+1)-\tfrac{3}{4}} \\
	-B_v\sqrt{J(J+1)-\tfrac{3}{4}}			& T_v - \tfrac{A_v}{2}+B_v\left[J(J+1)+\tfrac{1}{4}\right]
	\end{pmatrix}
\end{equation}
and
\begin{equation}
	H(^4\Pi) = \left(\begin{smallmatrix}
	T_v + \tfrac{3A_v}{2}+B_v\left[J(J+1)-\tfrac{19}{4}\right] & -\sqrt{3}B_v\sqrt{J(J+1)-\tfrac{15}{4}} & 0 & 0 \\
	-\sqrt{3}B_v\sqrt{J(J+1)-\tfrac{15}{4}} & T_v + \tfrac{A_v}{2} + B_v\left[J(J+1)+\tfrac{5}{4}\right] & -2B_v\sqrt{J(J+1)-\tfrac{3}{4}} & 0 \\
	0 & -2B_v\sqrt{J(J+1)-\tfrac{3}{4}} & T_v - \tfrac{A_v}{2} + B_v \left[J(J+1)+\tfrac{13}{4}\right] & -\sqrt{3}B_v(J+\tfrac{1}{2}) \\
	0 & 0 & -\sqrt{3}B_v(J+\tfrac{1}{2}) & T_v - \tfrac{3A_v}{2}+B_v\left[J(J+1)+\tfrac{5}{4}\right]
	\end{smallmatrix}\right)	.
\end{equation}
The top-left component is the one with $\Omega = 3/2$ and $5/2$, respectively.

The parameters used in these Hamiltonians are listed in Table~\ref{tab:coef}. Although some identified transitions occur at fairly high $J$, the contributions of the $D_v$ coefficients are not important at the few-cm$^{-1}$ scale. The $D_v$ coefficients for the $|a\,^4\Pi_u, v^\p=0, 1\rangle$ states are $5.0455(189)\times10^{-6}~{\rm cm}^{-1}$ and $5.0567(176)\times10^{-6}~{\rm cm}^{-1}$, respectively~\cite{ShansenJMS1983}. We extrapolate the $D_v$ coefficients of the $X\,^2\Pi_g$ state from merged parameters in ref.~\cite{ScoxonJMS1984} to obtain $D_{21} = 5.43(85)\times10^{-6}~{\rm cm}^{-1}$ and $D_{22} = 5.34(92)\times10^{-6}~{\rm cm}^{-1}$. Even at the higher $J$'s, the contributions from $D_v$ cancel to be consistent with zero with uncertainties of a few times $0.1~{\rm cm}^{-1}$.

\begingroup
\begin{table}[h]
\caption{Coefficients used in energy calculations. Uncertainties are shown in parentheses.}\label{tab:coef}
\begin{ruledtabular}
\begin{tabular}{clll}
state								& $T_v / {\rm cm}^{-1}$ & $A_v / {\rm cm}^{-1}$ & $B_v / {\rm cm}^{-1}$ \\
\hline
$X\,^2\Pi_g, v=21$	&	$129896.9(2.0)$\footnotemark[1] & $177.0(1.0)$\footnotemark[2]
		& 1.25(3)\footnotemark[2] \\ 
$X\,^2\Pi_g, v=22$	& $131075(5)$\footnotemark[2]			& $174.0(1.0)$\footnotemark[2]
		& 1.25(1)\footnotemark[2] \\ 
$a\,^4\Pi_u, v=0$		& $129892(2)$\footnotemark[3] 		& $-47.7927(19)$\footnotemark[4] 
		& $1.096990(26)$\footnotemark[4] \\
$a\,^4\Pi_u, v=1$		& $130904(2)$\footnotemark[3]			& $-47.7997(21)$\footnotemark[4]
		& $1.081532(18)$\footnotemark[4]
\end{tabular}
\end{ruledtabular}
\footnotetext[1]{Ref.~\cite{SkongCJP1994}}
\footnotetext[2]{Ref.~\cite{SsongJCP1999}}
\footnotetext[3]{Ref.~\cite{SkongIJMSIP1996}}
\footnotetext[4]{Ref.~\cite{ShansenJMS1983}}
\end{table}
\endgroup

\newpage

\section{Tables of near-degeneracies}

Below are tables listing every pair of energy levels with $|\Delta E|<10~{\rm cm}^{-1}$ and $\Delta J = 0, \pm 1$. Also provided are the estimated linewidths and the eigenstate superposition coefficients from diagonalizing the effective Hamiltonians above. The uncertainties in $\Delta E$ are 3--6~${\rm cm}^{-1}$. They are highly correlated, however, such that even if these particular transitions are no longer within $10~{\rm cm}^{-1}$, others likely will be. By convention~\cite{SherzbergVolI1950,Sbrown_carrington}, the $F_i$ indicate the energy order of the eigenstates for a given $J$ with $F_1$ having the lowest energy. In the case (a) limit, the \OOp~$X\,^2\Pi_g$ state has $\Omega=\tfrac{1}{2}$ in $F_1$ and $\tfrac{3}{2}$ in $F_2$, while the $a\,^4\Pi_u$ state has $\Omega=\tfrac{5}{2}, \tfrac{3}{2}, \tfrac{1}{2}, -\tfrac{1}{2}$ in $F_{1,2,3,4}$.

\begin{longtable}{@{\extracolsep{\fill}}ccccrrrrrrrr}
\caption{The near-degeneracies $|X\,^2\Pi_g, v^\pp=21, J^\pp\rangle$ and $|a\,^4\Pi_u, v^\p=0, J^\p\rangle$, including the eigenstate superposition coefficients.}\\
\hline\hline
$X\,^2\Pi_g$ & $a\,^4\Pi_u$ & $J^\pp$ & $J^\p$ & $\Delta E / {\rm cm}^{-1}$ & $\frac{\Gamma}{2\pi} / {\rm Hz}$ & $c_{3/2}^{\pp}$ & $c_{1/2}^\pp$ & $c_{5/2}^\p$ & $c_{3/2}^\p$ & $c_{1/2}^\p$ & $c_{-1/2}^\p$ \Tstrut\Bstrut\\
\hline \endfirsthead
\caption{(continued)}\Tstrut\\
\hline
$X\,^2\Pi_g$ & $a\,^4\Pi_u$ & $J^\pp$ & $J^\p$ & $\Delta E / {\rm cm}^{-1}$ & $\frac{\Gamma}{2\pi} / {\rm Hz}$ & $c_{3/2}^{\pp}$ & $c_{1/2}^\pp$ & $c_{5/2}^\p$ & $c_{3/2}^\p$ & $c_{1/2}^\p$ & $c_{-1/2}^\p$ \Tstrut\Bstrut\\
\hline \endhead
$F_2$ & $F_4$ & $\frac{7}{2}$ & $\frac{9}{2}$ & $-8.88$ & $0.45$ & $-1.00$ & $0.03$ & $-0.00$ & $0.02$ & $-0.20$ & $0.98$ \Tstrut\\
$F_2$ & $F_4$ & $\frac{9}{2}$ & $\frac{11}{2}$ & $-7.30$ & $0.63$ & $-1.00$ & $0.04$ & $-0.00$ & $0.03$ & $-0.24$ & $0.97$ \Tstrut\\
$F_2$ & $F_4$ & $\frac{11}{2}$ & $\frac{13}{2}$ & $-5.91$ & $0.83$ & $-1.00$ & $0.04$ & $-0.00$ & $0.04$ & $-0.27$ & $0.96$ \Tstrut\\
$F_2$ & $F_4$ & $\frac{13}{2}$ & $\frac{15}{2}$ & $-4.71$ & $1.05$ & $-1.00$ & $0.05$ & $-0.01$ & $0.05$ & $-0.30$ & $0.95$ \Tstrut\\
$F_2$ & $F_4$ & $\frac{15}{2}$ & $\frac{17}{2}$ & $-3.71$ & $1.28$ & $-1.00$ & $0.06$ & $-0.01$ & $0.07$ & $-0.34$ & $0.94$ \Tstrut\\
$F_2$ & $F_4$ & $\frac{17}{2}$ & $\frac{19}{2}$ & $-2.91$ & $1.52$ & $-1.00$ & $0.06$ & $-0.01$ & $0.08$ & $-0.36$ & $0.93$ \Tstrut\\
$F_2$ & $F_4$ & $\frac{19}{2}$ & $\frac{21}{2}$ & $-2.32$ & $1.77$ & $-1.00$ & $0.07$ & $-0.01$ & $0.09$ & $-0.39$ & $0.92$ \Tstrut\\
$F_2$ & $F_4$ & $\frac{21}{2}$ & $\frac{23}{2}$ & $-1.95$ & $2.02$ & $-1.00$ & $0.08$ & $-0.01$ & $0.11$ & $-0.42$ & $0.90$ \Tstrut\\
$F_2$ & $F_4$ & $\frac{23}{2}$ & $\frac{25}{2}$ & $-1.80$ & $2.27$ & $-1.00$ & $0.08$ & $-0.02$ & $0.12$ & $-0.44$ & $0.89$ \Tstrut\\
$F_2$ & $F_4$ & $\frac{25}{2}$ & $\frac{27}{2}$ & $-1.88$ & $2.52$ & $-1.00$ & $0.09$ & $-0.02$ & $0.13$ & $-0.46$ & $0.88$ \Tstrut\\
$F_2$ & $F_4$ & $\frac{27}{2}$ & $\frac{29}{2}$ & $-2.19$ & $2.76$ & $-1.00$ & $0.10$ & $-0.03$ & $0.15$ & $-0.48$ & $0.86$ \Tstrut\\
$F_2$ & $F_4$ & $\frac{29}{2}$ & $\frac{31}{2}$ & $-2.74$ & $3.00$ & $-0.99$ & $0.11$ & $-0.03$ & $0.16$ & $-0.50$ & $0.85$ \Tstrut\\
$F_2$ & $F_4$ & $\frac{31}{2}$ & $\frac{33}{2}$ & $-3.54$ & $3.23$ & $-0.99$ & $0.11$ & $-0.03$ & $0.18$ & $-0.51$ & $0.84$ \Tstrut\\
$F_2$ & $F_4$ & $\frac{33}{2}$ & $\frac{35}{2}$ & $-4.60$ & $3.45$ & $-0.99$ & $0.12$ & $-0.04$ & $0.19$ & $-0.53$ & $0.83$ \Tstrut\\
$F_2$ & $F_4$ & $\frac{35}{2}$ & $\frac{37}{2}$ & $-5.90$ & $3.66$ & $-0.99$ & $0.13$ & $-0.04$ & $0.20$ & $-0.54$ & $0.81$ \Tstrut\\
$F_2$ & $F_4$ & $\frac{37}{2}$ & $\frac{39}{2}$ & $-7.47$ & $3.87$ & $-0.99$ & $0.13$ & $-0.05$ & $0.21$ & $-0.55$ & $0.80$ \Tstrut\\
$F_2$ & $F_4$ & $\frac{39}{2}$ & $\frac{41}{2}$ & $-9.30$ & $4.07$ & $-0.99$ & $0.14$ & $-0.05$ & $0.23$ & $-0.57$ & $0.79$ \Tstrut\Bstrut\\
\hline
$F_1$ & $F_1$ & $\frac{7}{2}$ & $\frac{5}{2}$ & $-3.90$ & $0.07$ & $-0.03$ & $-1.00$ & $1.00$ & $0.08$ & $0.00$ & $0.00$ \Tstrut\\
$F_1$ & $F_1$ & $\frac{9}{2}$ & $\frac{7}{2}$ & $-7.85$ & $0.16$ & $-0.04$ & $-1.00$ & $0.99$ & $0.12$ & $0.01$ & $0.00$ \Tstrut\Bstrut\\
\hline
$F_1$ & $F_1$ & $\frac{5}{2}$ & $\frac{5}{2}$ & $4.79$ & $0.07$ & $-0.02$ & $-1.00$ & $1.00$ & $0.08$ & $0.00$ & $0.00$ \Tstrut\\
$F_1$ & $F_1$ & $\frac{7}{2}$ & $\frac{7}{2}$ & $3.32$ & $0.16$ & $-0.03$ & $-1.00$ & $0.99$ & $0.12$ & $0.01$ & $0.00$ \Tstrut\\
$F_1$ & $F_1$ & $\frac{9}{2}$ & $\frac{9}{2}$ & $1.43$ & $0.27$ & $-0.04$ & $-1.00$ & $0.99$ & $0.16$ & $0.02$ & $0.00$ \Tstrut\\
$F_1$ & $F_1$ & $\frac{11}{2}$ & $\frac{11}{2}$ & $-0.87$ & $0.41$ & $-0.04$ & $-1.00$ & $0.98$ & $0.19$ & $0.02$ & $0.00$ \Tstrut\\
$F_1$ & $F_1$ & $\frac{13}{2}$ & $\frac{13}{2}$ & $-3.59$ & $0.57$ & $-0.05$ & $-1.00$ & $0.97$ & $0.23$ & $0.03$ & $0.00$ \Tstrut\\
$F_1$ & $F_1$ & $\frac{15}{2}$ & $\frac{15}{2}$ & $-6.71$ & $0.74$ & $-0.06$ & $-1.00$ & $0.97$ & $0.26$ & $0.04$ & $0.00$ \Tstrut\Bstrut\\
\hline
$F_1$ & $F_1$ & $\frac{17}{2}$ & $\frac{19}{2}$ & $9.42$ & $1.12$ & $-0.06$ & $-1.00$ & $0.95$ & $0.31$ & $0.06$ & $0.01$ \Tstrut\\
$F_1$ & $F_1$ & $\frac{19}{2}$ & $\frac{21}{2}$ & $7.57$ & $1.33$ & $-0.07$ & $-1.00$ & $0.94$ & $0.34$ & $0.08$ & $0.01$ \Tstrut\\
$F_1$ & $F_1$ & $\frac{21}{2}$ & $\frac{23}{2}$ & $5.34$ & $1.55$ & $-0.08$ & $-1.00$ & $0.93$ & $0.37$ & $0.09$ & $0.01$ \Tstrut\\
$F_1$ & $F_1$ & $\frac{23}{2}$ & $\frac{25}{2}$ & $2.72$ & $1.76$ & $-0.08$ & $-1.00$ & $0.92$ & $0.39$ & $0.10$ & $0.02$ \Tstrut\\
$F_1$ & $F_1$ & $\frac{25}{2}$ & $\frac{27}{2}$ & $-0.28$ & $1.99$ & $-0.09$ & $-1.00$ & $0.90$ & $0.41$ & $0.11$ & $0.02$ \Tstrut\\
$F_1$ & $F_1$ & $\frac{27}{2}$ & $\frac{29}{2}$ & $-3.67$ & $2.21$ & $-0.10$ & $-1.00$ & $0.89$ & $0.43$ & $0.13$ & $0.02$ \Tstrut\\
$F_1$ & $F_1$ & $\frac{29}{2}$ & $\frac{31}{2}$ & $-7.42$ & $2.43$ & $-0.11$ & $-0.99$ & $0.88$ & $0.45$ & $0.14$ & $0.03$ \Tstrut\Bstrut\\
\hline
$F_1$ & $F_2$ & $\frac{25}{2}$ & $\frac{23}{2}$ & $4.95$ & $9.38$ & $-0.09$ & $-1.00$ & $-0.36$ & $0.82$ & $0.44$ & $0.10$ \Tstrut\\
$F_1$ & $F_2$ & $\frac{27}{2}$ & $\frac{25}{2}$ & $-1.75$ & $9.16$ & $-0.10$ & $-1.00$ & $-0.39$ & $0.79$ & $0.46$ & $0.11$ \Tstrut\\
$F_1$ & $F_2$ & $\frac{29}{2}$ & $\frac{27}{2}$ & $-8.77$ & $8.93$ & $-0.11$ & $-0.99$ & $-0.41$ & $0.76$ & $0.48$ & $0.13$ \Tstrut\Bstrut\\
\hline
$F_1$ & $F_2$ & $\frac{35}{2}$ & $\frac{35}{2}$ & $5.91$ & $8.04$ & $-0.13$ & $-0.99$ & $-0.48$ & $0.66$ & $0.55$ & $0.18$ \Tstrut\\
$F_1$ & $F_2$ & $\frac{37}{2}$ & $\frac{37}{2}$ & $-0.21$ & $7.82$ & $-0.13$ & $-0.99$ & $-0.50$ & $0.63$ & $0.56$ & $0.19$ \Tstrut\\
$F_1$ & $F_2$ & $\frac{39}{2}$ & $\frac{39}{2}$ & $-6.63$ & $7.61$ & $-0.14$ & $-0.99$ & $0.51$ & $-0.60$ & $-0.58$ & $-0.21$ \Tstrut\Bstrut\\
\hline
$F_1$ & $F_2$ & $\frac{51}{2}$ & $\frac{53}{2}$ & $5.66$ & $6.33$ & $-0.18$ & $-0.98$ & $-0.58$ & $0.44$ & $0.62$ & $0.29$ \Tstrut\\
$F_1$ & $F_2$ & $\frac{53}{2}$ & $\frac{55}{2}$ & $-0.72$ & $6.18$ & $-0.18$ & $-0.98$ & $-0.59$ & $0.42$ & $0.62$ & $0.30$ \Tstrut\\
$F_1$ & $F_2$ & $\frac{55}{2}$ & $\frac{57}{2}$ & $-7.39$ & $6.04$ & $-0.19$ & $-0.98$ & $0.60$ & $-0.40$ & $-0.63$ & $-0.30$ \Tstrut\Bstrut\\
\hline
$F_1$ & $F_3$ & $\frac{41}{2}$ & $\frac{39}{2}$ & $8.89$ & $7.13$ & $-0.15$ & $-0.99$ & $-0.19$ & $0.57$ & $-0.57$ & $-0.56$ \Tstrut\\
$F_1$ & $F_3$ & $\frac{43}{2}$ & $\frac{41}{2}$ & $1.14$ & $6.94$ & $-0.15$ & $-0.99$ & $0.21$ & $-0.58$ & $0.55$ & $0.57$ \Tstrut\\
$F_1$ & $F_3$ & $\frac{45}{2}$ & $\frac{43}{2}$ & $-6.87$ & $6.76$ & $-0.16$ & $-0.99$ & $-0.22$ & $0.59$ & $-0.52$ & $-0.58$ \Tstrut\Bstrut\\
\hline
$F_1$ & $F_3$ & $\frac{57}{2}$ & $\frac{57}{2}$ & $2.66$ & $5.71$ & $-0.20$ & $-0.98$ & $0.29$ & $-0.62$ & $0.37$ & $0.63$ \Tstrut\\
$F_1$ & $F_3$ & $\frac{59}{2}$ & $\frac{59}{2}$ & $-5.06$ & $5.58$ & $-0.20$ & $-0.98$ & $-0.30$ & $0.62$ & $-0.35$ & $-0.63$ \Tstrut\Bstrut\\
\hline
$F_1$ & $F_3$ & $\frac{75}{2}$ & $\frac{77}{2}$ & $8.43$ & $4.74$ & $-0.25$ & $-0.97$ & $0.37$ & $-0.62$ & $0.21$ & $0.66$ \Tstrut\\
$F_1$ & $F_3$ & $\frac{77}{2}$ & $\frac{79}{2}$ & $0.36$ & $4.67$ & $-0.25$ & $-0.97$ & $-0.37$ & $0.62$ & $-0.20$ & $-0.66$ \Tstrut\\
$F_1$ & $F_3$ & $\frac{79}{2}$ & $\frac{81}{2}$ & $-7.99$ & $4.60$ & $-0.26$ & $-0.97$ & $0.38$ & $-0.62$ & $0.18$ & $0.66$ \Tstrut\Bstrut\\
\hline
$F_1$ & $F_4$ & $\frac{61}{2}$ & $\frac{59}{2}$ & $1.74$ & $5.45$ & $-0.21$ & $-0.98$ & $-0.09$ & $0.32$ & $-0.63$ & $0.70$ \Tstrut\\
$F_1$ & $F_4$ & $\frac{63}{2}$ & $\frac{61}{2}$ & $-6.89$ & $5.57$ & $-0.21$ & $-0.98$ & $0.09$ & $-0.33$ & $0.63$ & $-0.69$ \Tstrut\Bstrut\\
\hline
$F_1$ & $F_4$ & $\frac{79}{2}$ & $\frac{79}{2}$ & $3.89$ & $6.38$ & $-0.26$ & $-0.97$ & $0.13$ & $-0.39$ & $0.66$ & $-0.63$ \Tstrut\\
$F_1$ & $F_4$ & $\frac{81}{2}$ & $\frac{81}{2}$ & $-4.98$ & $6.45$ & $-0.26$ & $-0.96$ & $0.13$ & $-0.39$ & $0.66$ & $-0.63$ \Tstrut\Bstrut\\
\hline
$F_1$ & $F_4$ & $\frac{101}{2}$ & $\frac{103}{2}$ & $6.82$ & $7.01$ & $-0.31$ & $-0.95$ & $0.16$ & $-0.44$ & $0.67$ & $-0.58$ \Tstrut\\
$F_1$ & $F_4$ & $\frac{103}{2}$ & $\frac{105}{2}$ & $-2.97$ & $7.05$ & $-0.31$ & $-0.95$ & $0.17$ & $-0.44$ & $0.67$ & $-0.57$ \Tstrut\Bstrut\\
\hline\hline
\end{longtable}

\begin{longtable}{@{\extracolsep{\fill}}ccccrrrrrrrr}
\caption{The near-degeneracies $|X\,^2\Pi_g, v^\pp=22, J^\pp\rangle$ and $|a\,^4\Pi_u, v^\p=1, J^\p\rangle$, including the eigenstate superposition coefficients.}\\
\hline\hline
$X\,^2\Pi_g$ & $a\,^4\Pi_u$ & $J^\pp$ & $J^\p$ & $\Delta E / {\rm cm}^{-1}$ & $\frac{\Gamma}{2\pi} / {\rm Hz}$ & $c_{3/2}^{\pp}$ & $c_{1/2}^\pp$ & $c_{5/2}^\p$ & $c_{3/2}^\p$ & $c_{1/2}^\p$ & $c_{-1/2}^\p$ \Tstrut\Bstrut\\
\hline \endfirsthead
\caption{(continued)}\Tstrut\\
\hline
$X\,^2\Pi_g$ & $a\,^4\Pi_u$ & $J^\pp$ & $J^\p$ & $\Delta E / {\rm cm}^{-1}$ & $\frac{\Gamma}{2\pi} / {\rm Hz}$ & $c_{3/2}^{\pp}$ & $c_{1/2}^\pp$ & $c_{5/2}^\p$ & $c_{3/2}^\p$ & $c_{1/2}^\p$ & $c_{-1/2}^\p$ \Tstrut\Bstrut\\
\hline \endhead
$F_1$ & $F_4$ & $\frac{1}{2}$ & $\frac{3}{2}$ & $-7.84$ & $0.08$ & $0.00$ & $1.00$ & $0.00$ & $0.00$ & $-0.08$ & $1.00$ \Tstrut\\
$F_1$ & $F_4$ & $\frac{3}{2}$ & $\frac{5}{2}$ & $-5.77$ & $0.17$ & $-0.01$ & $-1.00$ & $-0.00$ & $0.01$ & $-0.12$ & $0.99$ \Tstrut\\
$F_1$ & $F_4$ & $\frac{5}{2}$ & $\frac{7}{2}$ & $-3.87$ & $0.30$ & $-0.02$ & $-1.00$ & $-0.00$ & $0.01$ & $-0.16$ & $0.99$ \Tstrut\\
$F_1$ & $F_4$ & $\frac{7}{2}$ & $\frac{9}{2}$ & $-2.15$ & $0.45$ & $-0.03$ & $-1.00$ & $-0.00$ & $0.02$ & $-0.20$ & $0.98$ \Tstrut\\
$F_1$ & $F_4$ & $\frac{9}{2}$ & $\frac{11}{2}$ & $-0.61$ & $0.63$ & $-0.04$ & $-1.00$ & $-0.00$ & $0.03$ & $-0.23$ & $0.97$ \Tstrut\\
$F_1$ & $F_4$ & $\frac{11}{2}$ & $\frac{13}{2}$ & $0.75$ & $0.84$ & $-0.04$ & $-1.00$ & $-0.00$ & $0.04$ & $-0.27$ & $0.96$ \Tstrut\\
$F_1$ & $F_4$ & $\frac{13}{2}$ & $\frac{15}{2}$ & $1.92$ & $1.06$ & $-0.05$ & $-1.00$ & $-0.00$ & $0.05$ & $-0.30$ & $0.95$ \Tstrut\\
$F_1$ & $F_4$ & $\frac{15}{2}$ & $\frac{17}{2}$ & $2.90$ & $1.30$ & $-0.06$ & $-1.00$ & $-0.01$ & $0.06$ & $-0.33$ & $0.94$ \Tstrut\\
$F_1$ & $F_4$ & $\frac{17}{2}$ & $\frac{19}{2}$ & $3.67$ & $1.55$ & $-0.06$ & $-1.00$ & $-0.01$ & $0.08$ & $-0.36$ & $0.93$ \Tstrut\\
$F_1$ & $F_4$ & $\frac{19}{2}$ & $\frac{21}{2}$ & $4.24$ & $1.80$ & $-0.07$ & $-1.00$ & $-0.01$ & $0.09$ & $-0.39$ & $0.92$ \Tstrut\\
$F_1$ & $F_4$ & $\frac{21}{2}$ & $\frac{23}{2}$ & $4.59$ & $2.06$ & $-0.08$ & $-1.00$ & $-0.01$ & $0.10$ & $-0.41$ & $0.91$ \Tstrut\\
$F_1$ & $F_4$ & $\frac{23}{2}$ & $\frac{25}{2}$ & $4.73$ & $2.31$ & $-0.09$ & $-1.00$ & $-0.02$ & $0.12$ & $-0.44$ & $0.89$ \Tstrut\\
$F_1$ & $F_4$ & $\frac{25}{2}$ & $\frac{27}{2}$ & $4.63$ & $2.57$ & $-0.09$ & $-1.00$ & $-0.02$ & $0.13$ & $-0.46$ & $0.88$ \Tstrut\\
$F_1$ & $F_4$ & $\frac{27}{2}$ & $\frac{29}{2}$ & $4.31$ & $2.82$ & $-0.10$ & $-0.99$ & $-0.02$ & $0.14$ & $-0.48$ & $0.87$ \Tstrut\\
$F_1$ & $F_4$ & $\frac{29}{2}$ & $\frac{31}{2}$ & $3.74$ & $3.06$ & $-0.11$ & $-0.99$ & $-0.03$ & $0.16$ & $-0.49$ & $0.85$ \Tstrut\\
$F_1$ & $F_4$ & $\frac{31}{2}$ & $\frac{33}{2}$ & $2.93$ & $3.30$ & $-0.11$ & $-0.99$ & $-0.03$ & $0.17$ & $-0.51$ & $0.84$ \Tstrut\\
$F_1$ & $F_4$ & $\frac{33}{2}$ & $\frac{35}{2}$ & $1.87$ & $3.53$ & $-0.12$ & $-0.99$ & $-0.04$ & $0.18$ & $-0.53$ & $0.83$ \Tstrut\\
$F_1$ & $F_4$ & $\frac{35}{2}$ & $\frac{37}{2}$ & $0.56$ & $3.75$ & $-0.13$ & $-0.99$ & $-0.04$ & $0.20$ & $-0.54$ & $0.82$ \Tstrut\\
$F_1$ & $F_4$ & $\frac{37}{2}$ & $\frac{39}{2}$ & $-1.01$ & $3.97$ & $-0.13$ & $-0.99$ & $0.04$ & $-0.21$ & $0.55$ & $-0.81$ \Tstrut\\
$F_1$ & $F_4$ & $\frac{39}{2}$ & $\frac{41}{2}$ & $-2.85$ & $4.17$ & $-0.14$ & $-0.99$ & $0.05$ & $-0.22$ & $0.56$ & $-0.79$ \Tstrut\\
$F_1$ & $F_4$ & $\frac{41}{2}$ & $\frac{43}{2}$ & $-4.95$ & $4.36$ & $-0.15$ & $-0.99$ & $-0.05$ & $0.23$ & $-0.57$ & $0.78$ \Tstrut\\
$F_1$ & $F_4$ & $\frac{43}{2}$ & $\frac{45}{2}$ & $-7.32$ & $4.55$ & $-0.15$ & $-0.99$ & $-0.06$ & $0.25$ & $-0.58$ & $0.77$ \Tstrut\\
$F_1$ & $F_4$ & $\frac{45}{2}$ & $\frac{47}{2}$ & $-9.97$ & $4.73$ & $-0.16$ & $-0.99$ & $-0.06$ & $0.26$ & $-0.59$ & $0.76$ \Tstrut\Bstrut\\
\hline\hline
\end{longtable}

\section{Other degeneracies}

\begin{figure}[h]
	\includegraphics[width=0.75\columnwidth]{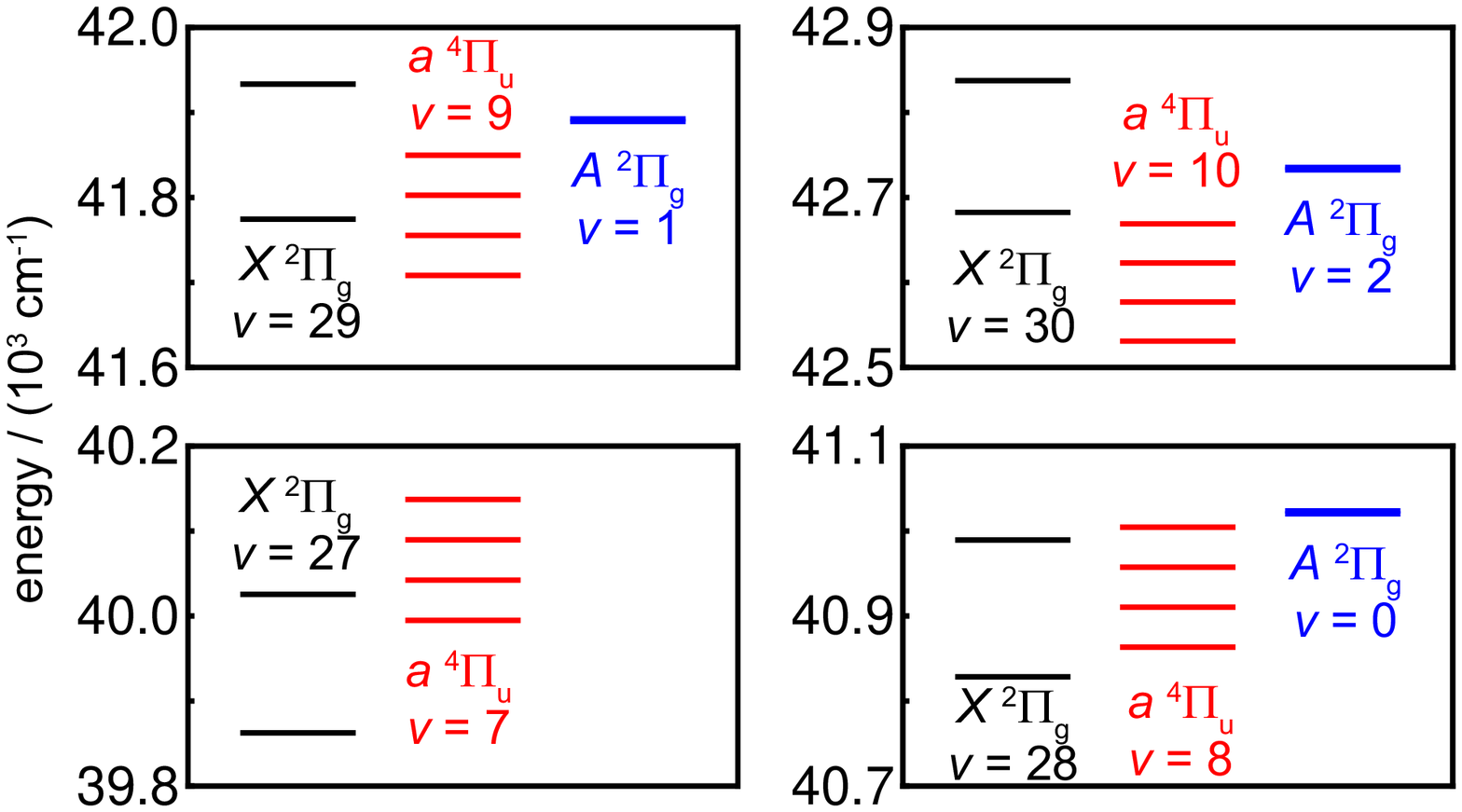}%
	\caption{Overlap of the $X\,^2\Pi_g$, $a\,^4\Pi_u$, and $A\,^2\Pi_u$ states near $|X, v=27\textnormal{--}30\rangle$. The levels are calculated from refs.~\cite{SsongJCP1999,SsongJCP2000,SsongJCP2000a}. Note that the $A$ state's spin-orbit constant is small enough that the doublet splitting is not visible at this scale.\label{fig:moreDegen}}
\end{figure}

\end{document}